\documentclass[twocolumn,aps,prl,showpacs]{revtex4}
\usepackage{graphicx}

\begin{document}

\title{Pattern scaling in the axial segregation of granular materials in a rotating tube}
\author{Christopher R. J. Charles, Zeina S. Khan and Stephen W. Morris}
\affiliation{Department of Physics, University of Toronto, 60 St. George St., 
 Toronto, Ontario, Canada M5S 1A7 }
\date{\today}


\begin{abstract}
Granular mixtures frequently segregate by grain size along the axis of partially-filled, horizontal, rotating tubes. When segregation approaches saturation at the surface, a well-defined pattern of bands with wavelength $\lambda$ emerges. The long-term dynamics of the pattern involves a much slower coarsening process. We characterized the initial saturated wavelength $\lambda$ as a function of the diameter of the tube $D$ for a filling fraction of 30 $\%$, for $1.9~{\rm cm} \leq D \leq  11.5~{\rm cm}$. We also studied the initial growth-rate of the bands as $D$ varies. We find that $\lambda/D$ is not constant, but rather increases rapidly for small $D$.  The growth-rate of bands decreases with smaller $D$ and segregation is suppressed completely for sufficiently small $D$.  These relatively simple features are not captured by any of the existing models of axial segregation.

~~

PREPRINT will be submitted to {\it Phys. Rev. E.} brief reports
\end{abstract}
\maketitle

\begin{figure}
\includegraphics[width=3.2in]{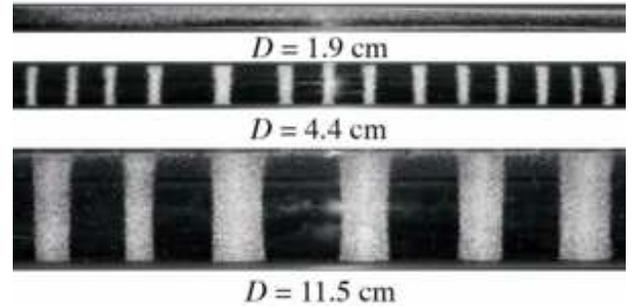}
\caption{Three tubes of different diameters, showing segregated bands. No clear segregation was seen for $D < 2.35$~cm. }
\label{photo_of_bands}
\end{figure}

A peculiar phenomenon occurs when two granular species with similar densities but different grain sizes are mixed together in long partially-filled, horizontal rotating tube. Rather than mixing together as one might expect, the grains segregate partially or completely by size \cite{Oyama, Duranspg,ristow}. Two distinct segregation phenomena have been observed, termed {\it radial}
 and {\it axial}. In radial segregation, the smaller component moves toward the axis of rotation and forms a buried core. The core typically threads the entire tube and, depending on the rotation speed, forms rather quickly, often within the first few tube revolutions \cite{KHprl,KHspg}.  The buried core is not usually visible from the surface. Radial segregation is followed by slower axial segregation in which the mixture separates into bands arranged along the axis of the tube as shown in figure \ref{photo_of_bands}.  The array of axial bands usually forms after several hundred to several thousand tube rotations.  It exhibits a reasonably well-defined and quasistable emergent wavelength $\lambda$.  The Fourier spectrum of the band pattern prior to saturation reveals a strong low-$\lambda$ cutoff, a sharp peak with a long tail to higher wavelengths (see Fig. 5 of Ref.~\cite{KCpre}). After a very long time, further rotation yields either complete segregation \cite{chicharro} or a smaller number of bands produced by coarsening effects\cite{ottino_prl,losert,KHpre2}.  

An obvious question of pattern scaling presents itself.  How does the emergent wavelength $\lambda$ scale with the other simple length scales in the system, the tube diameter $D$ and the average grain size $d$?  Although the phenomenon of axial segregation has been the subject of numerous studies\cite{Oyama,KHprl,KHspg, KCpre,chicharro,ottino_prl,losert,KHpre2,KCprl,zeina, Ziketal, Levinechaos,Levitan, elperin,ATprl,ATVpre,ams_gm}, its scaling behavior has not been systematically explored.  A recent study\cite{ams_gm} of the existence of axial segregation as a function of the dimensionless parameter $\delta = D/d$ suggests that a scaling approach could be fruitful. In this paper, we investigate the scaling of $\lambda$ as a function of $D$, for fixed $d$.  Other parameters that might be relevant, such as the rotation frequency $\omega$, the filled volume fraction $\eta$ and the size composition fraction $\phi$, were held fixed\cite{KCprl, KCpre}.  The scaling behavior we observe, although very simple, is nevertheless a strong test of models of this effect\cite{Ziketal,Levinechaos,Levitan,elperin,ATprl,ATVpre}.  

\begin{figure}
\includegraphics[width=3in]{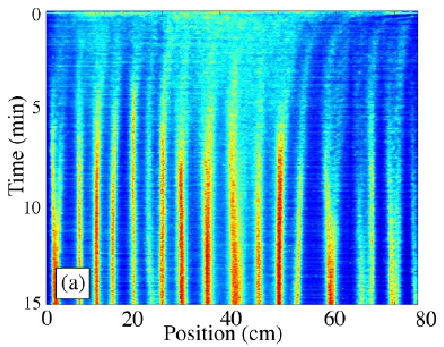}
\includegraphics[width=3in]{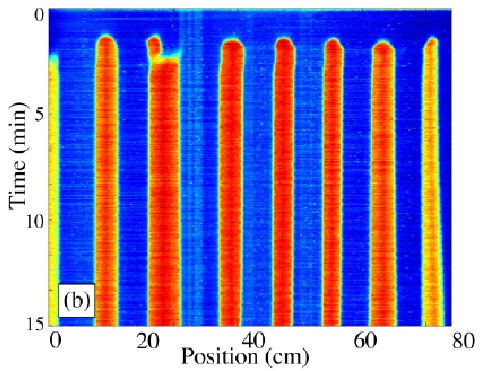}
\caption{Spacetime images of (a) $D$ = 2.85 cm and (b) $D$ = 11.5 cm showing the emergent wavelength $\lambda$. Images are false-coloured with red corresponding to the light colored grains. }
\label{xt_images}
\end{figure}

Nine glass tubes 1 m long with diameters $1.9 ~{\rm cm} \leq D \leq  11.5~ {\rm cm}$ were rotated at a constant rate of $\omega = 2.5~ {\rm rad}/{\rm s}$.  The grains consisted of glass spheres.  The larger (black) grains were 750 $\mu$m in diameter
%
%
 while the smaller (transparent) grains had diameters in the range 297 - 420 $\mu$m.  Thus, $d = 550 \mu$m.  The tubes were filled with equal volumes of each size, so that $\phi=0.5$, to a filled volume fraction $\eta =0.30$.  The grains were randomly pre-mixed and loaded into the tubes using a U-shaped channel \cite{KCprl,KCpre,zeina}. Static charging effects were reduced by adding a small quantity of Larostat FPE-S antistatic powder\cite{larostat}. The tube was illuminated from above while a CCD camera viewed the grains perpendicular to the flowing surface. Spacetime images like those shown in Fig.~\ref{xt_images} were constructed from horizontal strips 1 pixel high and 640 pixels long, stacked at 2 s or 3 s intervals. The horizontal strip captured the central 79 cm of the tube.
 
 For each $D$, the number of axial bands was counted after 5 min and  averaged over 6 or 7 runs. We thus obtaining the average wavelength $\lambda$ for each tube diameter $D$.  To investigate the growth rate of the axial pattern, we examined the power in the largest Fourier peak as a function of time.  These were found to increase approximately exponentially toward saturation\cite{KCpre}.  We fit these to extract the growth rate, which was a strong function of $D$, as discussed below.

%

Axial segregation was observed for tube diameters in the range $2.85~{\rm cm} \leq D \leq 11.5 ~{\rm cm}$. No segregation was observed for $D = 1.9 ~{\rm cm}$ even after several hours of rotation. For $D = 2.35 ~{\rm cm}$, segregation was irregular and incomplete for short times (less than 2 hours) but long-term segregation was observed after 15 hours of rotation.
	
\begin{figure}
\includegraphics[width=3.2in]{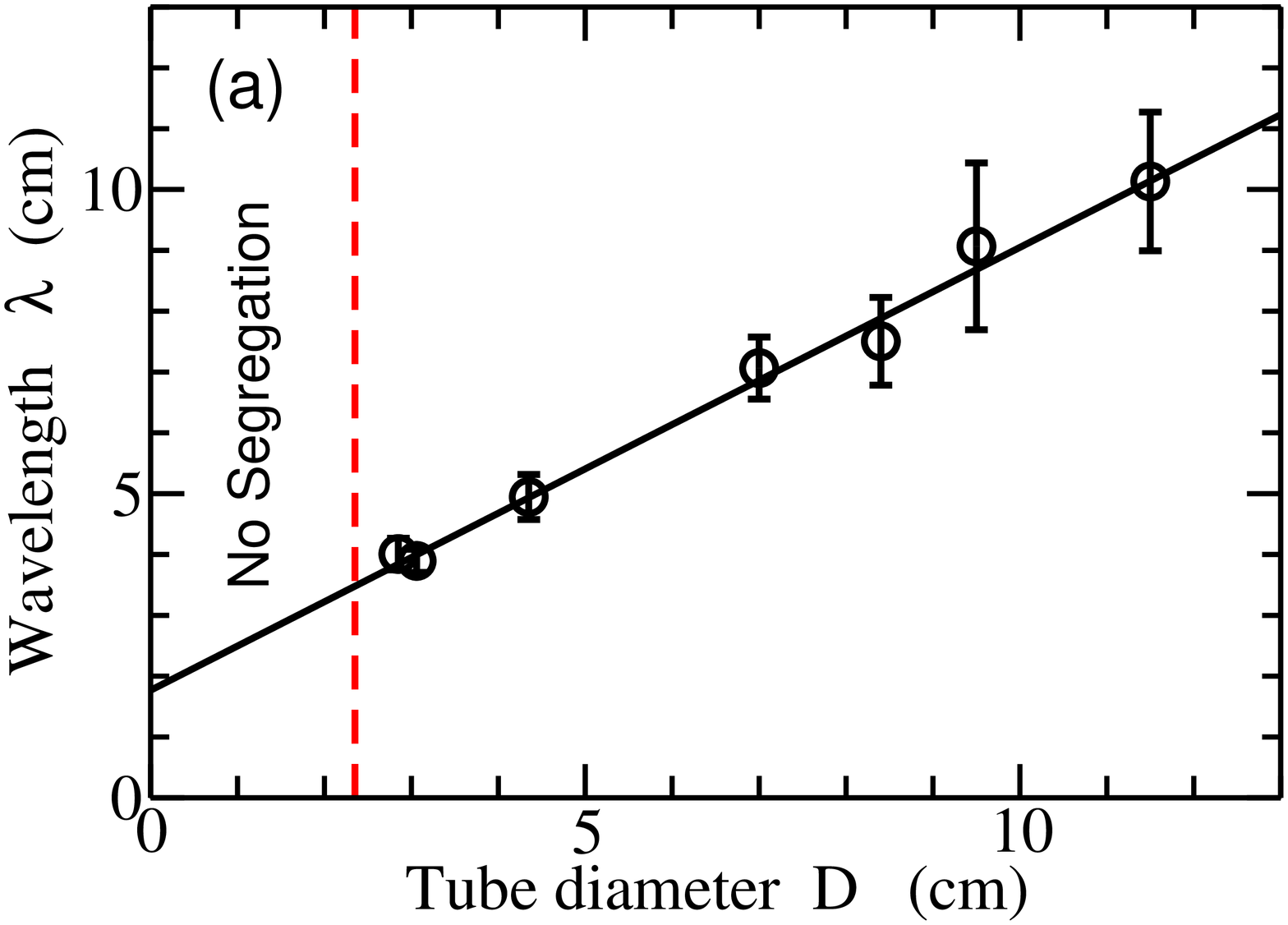}
\includegraphics[width=3.2in]{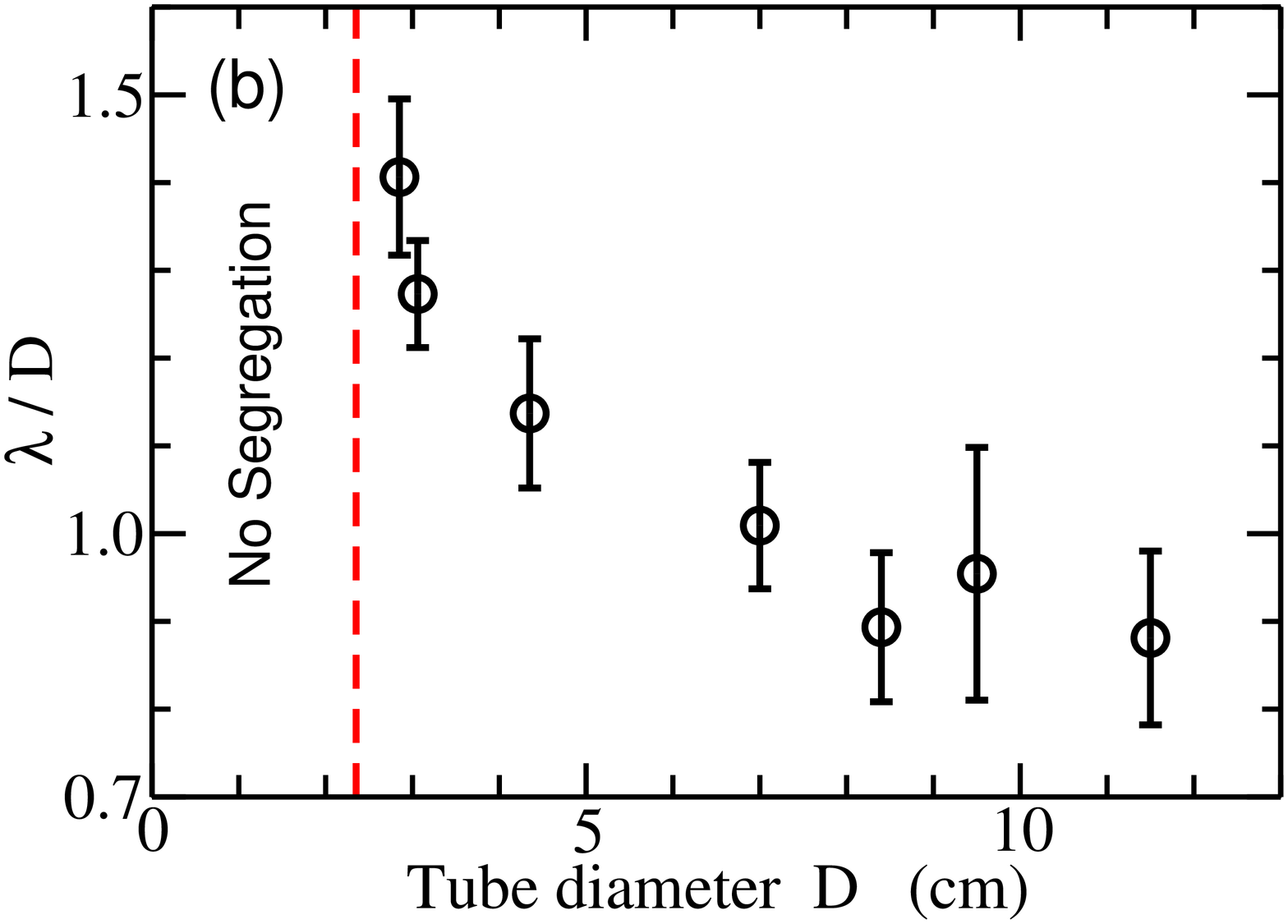}
\caption{Axial segregation results showing (a) $\lambda$ vs. $D$ and (b) $\lambda/D$ vs. $D$. No segregation is observed for tubes smaller than $D=$2.35~cm}
\label{lam_vs_D}
\end{figure}

\begin{figure}
\includegraphics[width=2.8in]{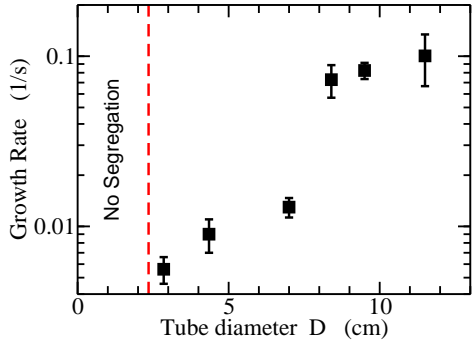}
\caption{Growth rates for six tubes of different diameters.  No segregation, and hence a zero growth rate, is found for tubes with $D$ smaller than 2.35~cm. }
\label{growth_vs_D}
\end{figure}

The averaged wavelength $\lambda$ for each $D$ is plotted in Fig.~\ref{lam_vs_D}(a).  A linear fit does not intercept the origin, indicating that $\lambda$ is not directly proportional to $D$.  The slope is $0.73\pm 0.03$, while the intercept is  $1.8\pm 0.2~ {\rm cm}$. Figure \ref{lam_vs_D}(b) shows the ratio $\lambda/D$ {\it vs}. $D$ from the same data. For small $D$, $\lambda/D$ increases toward the region where segregation is not observed.  For large D, $\lambda/D$ tends toward a constant equal to the slope from  Fig.~\ref{lam_vs_D}(a).  Turning to the growth rates, Fig.~\ref{growth_vs_D} shows the results for the six tube diameters where axial segregation occurred. The growth rate of axially segregated bands is a rapidly increasing function of $D$.
	
Our results for $1.9~ {\rm cm} \leq D \leq 11.5~ {\rm cm}$ are broadly consistent with the empirical survey presented in Ref.~\cite{ams_gm}. That study concluded that if the ratio $\delta = d/D$ is larger than a critical value $\delta_c \approx 55$, axial bands always form, while if $\delta < 40$ segregation is never found. For our smallest tube, $D=1.9$~cm, in which no segregation was observed, $\delta = 35$. For $D = 2.35 {\rm cm}$, we find $\delta = 43$ and segregation was incomplete even after 2 hours of rotation.  For $D = 2.85 {\rm cm}$, $\delta = 51 \sim \delta_c$, but we found clear, if rather slow, axial segregation.  All other values of $D$ have $\delta > \delta_c$ and clear axial segregation was observed.

Continuum models of axial segregation\cite{Ziketal,Levinechaos,Levitan,elperin,ATprl,ATVpre} do not appear to make any useful predictions for this scaling.  In general, the models are scaled so that all lengths are expressed in units of $D$, and hence for constant parameters they apparently predict a constant $\lambda/D$.  In practice, the parameters of the various models probably contain some $D$ dependence, which is however unanalyzed within the models.  Some clue about this dependence could be gleaned from the strong $D$ dependence of the growth rate of the axial bands.  Continuum models typically do not explicitly contain any scaling with the mean particle size $d$, and hence cannot account for the dependence of segregation on the large parameter $\delta$.


In conclusion, we have examined the scaling of the pattern wavelength $\lambda$ with the tube diameter $D$. We find that $\lambda/D$ approaches a constant for large $D$, but increases significantly as one approaches the region where the growth rate falls to zero, which ocurrs at small, but nonzero, $D$.  Our results generally agree with a recent study\cite{ams_gm} that found that segregation does not occur unless $\delta =D/d > 55$.   No model of which we are aware correctly predicts the scaling of the pattern with either the tube diameter $D$, or the mean grain size $d$. Thus, even the simplest scaling features of this striking effect remain unaccounted for.

\acknowledgments
We wish to thank Zahir Daya and Wayne Tokaruk. This work was supported by the Natural Science and Engineering Research Council of Canada.

\end{document}